\documentclass[twocolumn,superscriptaddress,preprintnumbers,amsmath,amssymb]{revtex4}
\usepackage{graphicx}
\usepackage{dcolumn}
\usepackage{bm}
\usepackage{color}
\usepackage{dcolumn}

\makeatletter
\@ifundefined{textcolor}{}
{%
 \definecolor{BLACK}{gray}{0}
 \definecolor{WHITE}{gray}{1}
 \definecolor{RED}{rgb}{1,0,0}
 \definecolor{GREEN}{rgb}{0,1,0}
 \definecolor{BLUE}{rgb}{0,0,1}
 \definecolor{CYAN}{cmyk}{1,0,0,0}
 \definecolor{MAGENTA}{cmyk}{0,1,0,0}
 \definecolor{YELLOW}{cmyk}{0,0,1,0}
}

\usepackage{dcolumn}
\usepackage{bm}
\usepackage[pdfstartview=FitH, CJKbookmarks=true, bookmarksnumbered=true, bookmarksopen=true, colorlinks=true, pdfborder=001, citecolor=blue, linkcolor=blue, urlcolor=blue, linktocpage=true] {hyperref}

\begin{document}

\title{Tunable multiphoton bundles emission in a Kerr-type two-photon Jaynes-Cummings model}

\author{Jing Tang}
\affiliation{School of Physics and Optoelectronic Engineering, Guangdong University of Technology, Guangzhou 510006, China}
\affiliation{Guangdong Provincial Key Laboratory of Sensing Physics and System Integration Applications, Guangdong University of Technology, Guangzhou, 510006, China}

\author{Yuangang Deng}
\email{dengyg3@mail.sysu.edu.cn}
\affiliation{Guangdong Provincial Key Laboratory of Quantum Metrology and Sensing $\&$ School of Physics and Astronomy, Sun Yat-Sen University (Zhuhai Campus), Zhuhai 519082, China}

\date{\today}

\begin{abstract}
We present a study on manipulation and enhancement of multiphoton bundles emission under a moderate atom-cavity coupling, by constructing a two-photon Jaynes-Cummings model integrated with Kerr nonlinearity in a single atom-cavity system.  We show that the vacuum-Rabi splittings for the $n$th dressed states can be significantly enhanced by Kerr interaction. This remarkable enhancement in energy-spectrum anharmonicity with the well-resolved multiphoton resonance facilitates the generation of special nonclassical states beyond the strong-coupling limit in the experiment. In particular, both two- and three-photon blockades are observed with adjusting the amplitude of the cavity-driving or atom-pump fields.  Moreover, we discover that transitions between two- and three-photon bundles can be achieved through tuning  the atom-cavity detuning or Kerr nonlinearity. It further showcases the three-photon blockade is substantially strengthened when both the cavity and atomic fields are jointly driven. Our proposal unveils a pathway  for realizing highly controllable nonclassical states and quantum devices with combining two-photon Jaynes-Cummings interactions and Kerr nonlinearity, which may pave the way for versatile applications in quantum information science, e.g., all-optical switches and transistors.
\end{abstract}

\maketitle
\section{Introduction}

The realization and manipulation of nonclassical photon emission are interesting and important research topics in modern science, with versatile application prospects ranging from quantum communication~\cite{Duan01, Scarani09}, to quantum metrology ~\cite{PhysRevLett.96.010401, Giovannetti2011}, and quantum computations~\cite{Knill01, Pieter07, Buluta11}.  Generally, generating nonclassical lights at the single- or few-photon levels requires strong photon-photon interactions. Photon blockade (PB) plays a crucial role in generating nonclassical light, characterized by sub-Poisson photon number statistics and antibunching with respect to single photons~\cite{Birnbaum2005}.  PB has been extensively studied both theoretically and experimentally in various quantum systems~\cite{Hennessy07, Fink08, PhysRevX.5.031028, Reinhard2012, Kai15,Tang21,Mirza17,  Gheeraert18,  Liu14, Rabl11, Nunnenkamp11,Wang15, PhysRevA.105.043711}. These pioneering advancements have revealed conventional photon blockade (CPB) based on strong energy spectrum anharmonicity~\cite{Dayan08, Faraon08,Hoffman11,Lang11}, and unconventional photon blockade (UPB) arising from quantum destructive interference between multiple pathways~\cite{Tang15, Flayac17, Tang19, Liew10, Majumdar2012, PhysRevLett.127.240402, Bamba2011, PhysRevA.89.031803, PhysRevA.92.023838, Sarma18, 2020Conventional, Snijders18, Vaneph18}.

Meanwhile, the nonclassical $n$-photon states via high-order transition processes have garnered considerable research interest, offering new resources for applications in multiphoton quantum nonlinear optics, such as multiphoton sources~\cite{Chang14,Chang16} and entangled photon sources~\cite{PhysRevLett.129.193604, Dousse2010, Muller2014,Yao12}. Particularly, two-photon blockade (two-PB) has been experimentally realized in the atom-cavity QED systems at strong coupling regime~\cite{PhysRevLett.118.133604}. Recent theoretical advancements have proposed the generation of multiphoton blockade in cavity QED systems through various driven configurations ~\cite{SHAMAILOV2010766, PhysRevA.95.063842, PhysRevA.98.043858, PhysRevA.102.053710, PhysRevA.104.053718, PhysRevA.99.053850}, such as qubit-cavity systems~\cite{PhysRevA.91.043831}, driven nonlinear systems~\cite{PhysRevA.87.023809,PhysRevA.90.013839, PhysRevA.100.053857} and cavity optomechanics based on squeezing~\cite{PhysRevA.103.043509} or mechanical driving~\cite{Zhai:19}. Additionally, $n$-quanta bundles emission~\cite{munoz2014emitters}, where photons (phonons) are released in groups of  strongly correlated $n$ photons (phonons), has been explored in superconducting microwave resonator or cavity QEDs~\cite{PhysRevA.93.013808, PhysRevResearch.3.043020, cosacchi2022nphoton,munoz2018filtering}, $n$-quanta Jaynes-Cummings (JC) type interaction~\cite{Deng:21,Tang:23, Jiang_2023}, Stokes or parity-symmetry-protected processes~\cite{PhysRevLett.124.053601,PhysRevLett.127.073602}, and high-spin atom-cavity system~\cite{tang2022strong}. These advancements highlight the requirement for sufficiently strong nonlinearities to engineer quantum light at the $n$-photon level. However, producing high-quality nonclassical multiphoton states remains challenging because of the inherently weak $n$-photon interactions~\cite{Schuster2008, xu2018}. Thus the new mechanism for engineering special nonclassical states beyond the strong coupling regime is desirable to study, with prospects for exploring versatile applications in quantum information and quantum networks.

 \begin{figure}[ht]
\includegraphics[width=0.96\columnwidth]{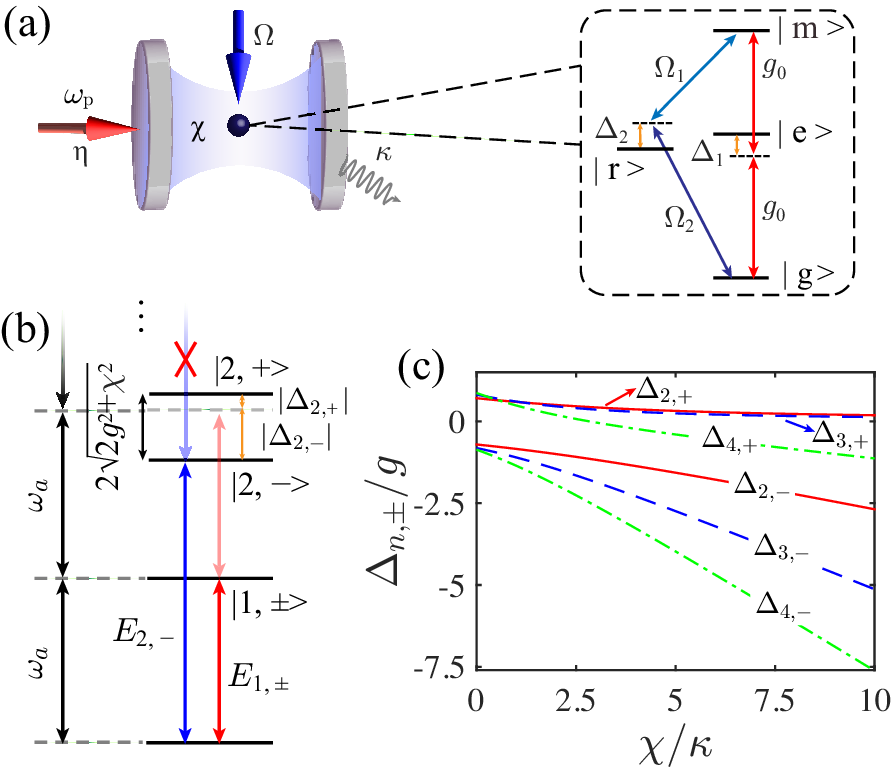}
\caption{(a) Scheme for generating multiphoton bundles emission in a single atom-cavity system with Kerr nonlinearity medium. The level diagram of the single atom is shown within the dashed box. (b) The typical energy spectrum of the system in the absence of the weak driven fields and dissipations of system. (c) Vacuum Rabi splitting $\Delta_{n,\pm}$ for the $n$th dressed states  vs Kerr nonlinearity $\chi$ with $n=2, 3, 4$. The parameters are $\Delta_a=2\Delta_c$, $\eta=\Omega=0$, $g/\kappa=4$, and $\kappa=2\pi \times 150$ kHz.} \label{scheme}
\end{figure}%

In this paper, we propose a scheme for the generation and manipulation of nonclassical light emission through the incorporation of Kerr nonlinearity in a two-photon Jaynes-Cummings framework for a moderate atom-cavity coupling.  By leveraging the interplay of two-photon JC type interaction and Kerr nonlinearity, we show that the vacuum Rabi splitting for the $n$th manifold is notably amplified with increasing Kerr nonlinearity via examining the energy spectrum analytically. This significant enhancement in energy-spectrum anharmonicity with the well-resolved multiphoton resonance, enables the versatility in generating highly controllable nonclassical light states such as strong single-PB, photon-induced tunneling (PIT), two-, and three-photon bundles, marking a significant advance in the field of quantum optics and photonics. The study delineates several significant findings: {\romannumeral1}) In the cavity-driven case, both three-PB and two-PB can be observed and interconverted by adjusting the cavity-driven strength, which is distinct from the PIT effect observed in the absence of Kerr nonlinearity~\cite{Tang:23}. Particularly, the system can function as a transducer between three-PB (or two-PB) and single-PB by tuning the cavity-light detuning. {\romannumeral2}) When driven by an atom-pump field, the transition between two- and three-photon bundles emission are achieved by tuning either the atom-pump strength or Kerr nonlinearity. {\romannumeral3}) The combined application of both cavity-driven and atom-pump fields reveals a highly tunable transducer among two-PB, PB, PIT, and three-PB, modulated by the cavity-light detuning. Remarkably, the three-PB at the red detuning of the two-photon resonance is significantly enhanced by controlling the atom-pump strength. Our investigation highlights the potential for generating quantum devices and fundamental phenomena within many-body physics~\cite{Chang14}, offering promising prospects for versatile applications in quantum information science and technology. 

This paper is organized as follows: Section \uppercase\expandafter{\romannumeral2} outlines the proposed scheme for generating a Kerr-type two-photon JC model and elaborates on the Hamiltonian of the system. Section \uppercase\expandafter{\romannumeral3} focuses on the examination of nonclassical multiphoton states. The paper concludes with a summary presented in Sec. \uppercase\expandafter{\romannumeral4}.

\section{Model and Hamiltonian}
We explore the emission of multiphoton bundles from a single four-level atom trapped inside a high-finesse cavity with Kerr nonlinearity. As illustrated  in  Fig~\ref{scheme}(a), a single atom is coupled to a single-mode cavity embedded in a Kerr nonlinearity medium with frequency $\omega_c$. The energy levels of the atom comprise the electronic ground state $|g\rangle$, two intermediate excited states $|e \rangle$, $|r \rangle$, and a long-lived electronic orbital state $|m\rangle$. The cavity mode couples to the atomic transitions $|g\rangle  \leftrightarrow |e\rangle$ and $|e\rangle  \leftrightarrow |m\rangle$ with a single atom-cavity coupling $g_0$ and an atom-cavity detuning $\Delta_1$. To generate an effective two-photon pump field, two laser beams with Rabi frequencies $\Omega_1$ and $\Omega_2$ are  employed to drive the atom transitions $|r\rangle  \leftrightarrow |m\rangle$ and $|g\rangle  \leftrightarrow |r\rangle$ in the far-resonant regime, characterized by the atom-pump detuning $\Delta_2$. Additionally, the optical cavity with a decay rate $\kappa$, is driven by a laser field with a single-photon parametric amplitude $\eta$ and frequency $\omega_p$.

In the rotating frame, the atom-cavity Hamiltonian under the rotating-wave approximation is presented as:
 \begin{align}\label{Hamiltonian}
 \mathcal {\hat{H}}_0/\hbar&= \Delta_c' \hat{a}^\dag\hat{a} +\Delta_1 \hat \sigma_{ ee}  +\Delta_2  \hat \sigma_{rr} +\delta_m  \hat \sigma_{mm}+ \chi  \hat{a}^\dag  \hat{a}^\dag\hat{a} \hat{a} \nonumber \\
&+ g_0 (\hat{a}^\dag  \hat \sigma_{ge}+ \hat{a}^\dag \hat \sigma_{ em})+g_0 (\hat{a}  \hat \sigma_{eg}+ \hat{a} \hat \sigma_{ me}) 
 \nonumber \\
&+(\Omega_2 \hat \sigma_{ gr} + \Omega_1\hat \sigma_{ rm} +\eta\hat{a}^\dag +{\rm H.c.}),
\end{align}%
where $\hat{a}^\dag$ ($\hat{a}$) indicates the creation (annihilation) operator of the cavity mode, $\hat{\sigma}_{ij}=|i\rangle  \langle j|$ denotes the atomic operators for $i, j=(g, e, r, m)$, and $\chi$ represents the strength of the Kerr nonlinearity. The detunings are defined with $\Delta_c'=\omega_c-\omega_p$, $\Delta_1=\omega_e-\omega_p$,  $\Delta_2=\omega_r-\omega_p$ as  the single-photon detunings, $\delta_m=\omega_m-2\omega_p$ as  the two-photon detuning, and $\hbar \omega_i$ marks the transition energy from $|g\rangle  \leftrightarrow |i\rangle$.

In the far-off resonance regime, where $|g_0/\Delta_1|\ll1$ and $|\Omega_2/\Delta_2|\ll1$, two intermediate excited states $|e_0\rangle$ and $|e_1\rangle$ are adiabatically eliminated. Then Hamiltonian (\ref{Hamiltonian}) reduces to
\begin{align}\label{Hamiltonian1}
 \mathcal {\hat{H}}_1/\hbar&= \Delta_c \hat{a}^\dag\hat{a}+ \Delta_a \hat \sigma_{mm}+ \chi  \hat{a}^\dag  \hat{a}^\dag\hat{a} \hat{a}+  \eta (\hat{a}^\dag+\hat{a})  \nonumber \\
&+g(\hat{a}^{\dag2} \hat{\sigma}_{gm}+\hat{a}^2 \hat{\sigma}_{mg}) + \Omega(\hat{\sigma}_{gm}+\hat{\sigma}_{mg}),
\end{align}%
with the effective cavity-light detuning $\Delta_c= \Delta_c'-g_0^2/\Delta_1$ and atom-light detuning $\Delta_a=\delta_m +\Omega_2^2/\Delta_2 - \Omega_1^{2} /\Delta_2$ including the optical Stark shift of single atom. Clearly, this Hamiltonian delineates  a Kerr-type two-photon JC model with a coupling strength $g=-g_0^2/\Delta_1$, emphasizing the direct two-photon transition between $|g\rangle$ and $|m\rangle$. Additionally, the last term describes the direct two-photon atom-pump process with the coupling strength $\Omega=-\Omega_1 \Omega_2 /\Delta_2$.  Based on the ratio
between the atom-cavity coupling $g$ and the cavity decay rate $\kappa$, three major coupling regimes can be defined, the strong-coupling regime with $g/\kappa \gg 1$, the moderate regime with $g/\kappa \sim 1$,  and the weak-coupling regime with $g/\kappa \ll 1$.

As we will see below, by combining the Kerr nonlinearity and two-photon JC type interaction, nonclassical states ranging from sing-PB to two-photon bundles emission and three-photon bundles emission for a moderate atom-cavity coupling are realized with driving either the cavity or the atom. Notably, with the Kerr nonlinearity is considered, the proposal of generating strong multiphoton bundles emission is essential different from the standard two-photon JC model~\cite{Tang:23}.  Unlike the two-photon Rabi model, which exhibits discrete ${\cal Z}_4$-symmetry, it is important to note that the proposed Hamiltonian (\ref{Hamiltonian1}) preserves the continuous ${\cal U}(1)$ symmetry in the absence of system dissipation, categorizing it within the supersymmetry (SUSY) class as discussed in Ref.~\cite{maldonado2021underlying}.

To gain insights into the multiphoton emission, it is insightful to analyze the energy spectrum of the Kerr-type two-photon JC model. With neglecting the weak cavity driving and pump fields, we note that the total excitation number $\hat{N}=\hat{a}^\dag\hat{a} +2\hat{\sigma}_{mm}$ commutes with the Hamiltonian $\mathcal {\hat{H}}_1$, which can be treated as a conserved quantity in our proposal. As a consequence, the Hilbert spaces can be restricted to $|n, g \rangle$ and $|n-2, m \rangle$, with $n$ denoting the photon excitation number. By solving the Schr$\ddot{o}$dinger equation ${\hat{\cal H}}\Psi={\cal M}\Psi$, where $\Psi=[|n, g\rangle,|n-2,m\rangle]^T$,  the relevant matrix ${\mathcal {M}}$ can be readily obtained as 
\begin{align}\label{matrix}
{\mathcal {M}}\!=\left(\!\begin{array}{ccc}
n\Delta_c +n(n-1)\chi & \sqrt{n(n-1)}g\\
\sqrt{\!n(n-\!1)}g\! &(\!n\!-\!2\!)\Delta_c\!+\!\Delta_a\!+\!(\!n\!-\!2)(\!n-\!3)\chi \\
\end{array}\right)\!
\end{align}   

The eigenstates of the system for $n\ge2$ can be expressed as $E_{n, \pm} =( |n, g\rangle \pm |n-2, m\rangle)/\sqrt{2}$, where $+$ ($-$) denotes the upper (lower) branch of the $n$th dressed states. The corresponding energy eigenvalues under the resonance condition ($\Delta_a=2\Delta_c$) are obtained by diagonalizing the matrix in Eq.~(\ref{matrix}),
\begin{align} \label{Energy}
E_{n, \pm} &=n\Delta_c+(n^2-3n+3)\chi  \nonumber \\
&\pm\sqrt{(2n-3)^2\chi^2+{n(n-1)}g^2},
\end{align}%
for $n\ge2$. Remarkably, the vacuum Rabi splitting $\Delta_{n,\pm}$ between the $n$th dressed states is nonlinear and increases with the increasing $\chi$,  leading to a significant enhancement of nonlinearity and energy spectrum anharmonicity.   For the case of one photon excitation, the eigenvalues $E_{1, \pm} = \Delta_c$ remain independent of Kerr nonlinearity.

To clarify the physical mechanism behind multiphoton emission, we depict a typical energy spectrum of the Kerr-type two-photon JC model under the resonance condition, as shown in Fig.\ref{scheme}(b). It can be seen that the energy spectrum in the rotating frame exhibits an anharmonic ladder, which is conducive to the generation of nonclassical multiphoton states. Furthermore, we plot  the vacuum Rabi splitting $\Delta_{n,\pm}$ in the few-photon subspaces ($n=2, 3, 4$) as a function of Kerr nonlinearity $\chi$, as shown in Fig.~\ref{scheme}(c). Notably, with increasing Kerr nonlinearity, the energy splittings for the $n$th dressed state monotonically increase. Interestingly, $\Delta_{2, +} $ and $\Delta_{3, +}$ are nearly degenerate across the entire parameter range of $\chi$, leading to the transition between two-PB and three-PB by adjusting the driving strength, as will be demonstrated below.

Based on the energy spectrum depicted in Fig.\ref{scheme}(b), we anticipate that in the cavity-driven case (red lines), single-PB can be generated as a result of the anharmonic ladder induced by the splitting of the second dressed state with a weak laser field. Conversely, in the atom-pump case (blue lines) via a two-photon transition, the system may behave as a highly nonclassical source in the two-photon and three-photon regimes because of the nearly degenerate $\Delta_{2, +} $ and $\Delta_{3, +}$. We have checked that the energy spectrum can be slightly distorted by the weak cavity-driven and atom-pump fields. In fact, a large driven strength can slightly adjust the dressed states with a proper $\chi$, while the multiphoton excitation processes facilitate the generation of multi-PB. This offers significant opportunities for  generating various nonclassical light states and exploring potential applications in quantum information sciences.
         
\section{Numerical results}
Taking into account of the complete dissipations, the out-of-equilibrium dynamics of the proposed two-photon Kerr-type JC model can be characterized by solving master equation,
 \begin{equation}\label{master equation}%
{ \frac{d\rho}{dt}}= -i [\hat{H}, {\rho}] + \frac{\kappa}{2} \mathcal
{\cal{D}}[\hat{a}]\rho + \frac{\gamma}{2} \mathcal
{\cal{D}}[\hat{\sigma}_{gm}]\rho,
\end{equation}
where  $\rho$ represents the corresponding density matrix of the single atom-cavity system. $\mathcal {D}[\hat{o}]\rho=2\hat{o} {\rho} \hat{o}^\dag - \hat{o}^\dag \hat{o}{\rho} - {\rho}\hat{o}^\dag \hat{o}$ denotes the Lindblad-type dissipation, and $\gamma$ is the effective atomic decay rate. Consequently, the quantum  statistics and steady-state photon number can be directly evaluated by numerically solving Eq.~(\ref{master equation}) with ${d\rho_s}/{dt}= 0$.     

The performance of quantum light from isolated photons to $n$-photon bundles can be characterized using the generalized $k$th-order correlation function defined as ~\cite{munoz2014emitters} 
\begin{align} 
g_n^{(k)}(\tau_1,\ldots,\tau_n)=\frac{\left\langle \prod_{i=1}^k\left[\hat{a}^{\dagger }(\tau_i)\right]^n \prod_{i=1}^k\left[\hat{a}(\tau_i)\right]^n\right\rangle}{\prod_{i=1}^k\left\langle \left[\hat{a}^{\dagger }(\tau_i)\right]^n\left[\hat{a}(\tau_i)\right]^n \right\rangle},
\label{g220}%
\end{align} 
with $\tau_1\leq...\leq\tau_n$. It is noteworthy that Eq.~({\ref{g220}) reverts to the standard $k$th-order correlation function for isolated photons when $n=1$, i.e., $g_1^{(k)}(0)=\mathcal {\rm Tr} (\hat{a}^{\dagger k} \hat{a}^k \rho_s)/n_s^k$, which can be obtained by numerically solving the steady-state density matrix in Eq.~(\ref{master equation}) with the steady-state intracavity photon number $n_s$=Tr$(\hat{a}^\dag\hat{a}\rho_s)$.  

Next, we introduce the criteria for the related nonclassical photon emission. Single-PB is characterized by the conditions $g^{(2)}_1(0)<1$ and $g_1^{(2)}(0)<g_1^{(2)}(\tau)$, indicating sub-Poissonian photon-number statistics and photon antibunching. While the photon-number statistics conditions for $n$PB ($n\ge2$) are  $g_1^{(n)}(0) >1$ and $g_1^{(n+1)}(0) < 1$, signifying $n$-photon super-Poissonian, and $(n + 1)$-photon sub-Poissonian photon statistics~\cite{PhysRevLett.118.133604,PhysRevLett.121.153601}. In particular, for $n$-photon bundles emission, additional conditions of $g_1^{(2)}(0)>g_1^{(2)}(\tau)$ and  $g_n^{(2)}(0)<g_n^{(2)}(\tau)$ must be fulfilled to guarantee bunching of the bundle photons and antibunching between separated bundles of photons~\cite{munoz2014emitters, Chang16}. In addition, PIT is defined  according to the higher-order correlation function $g_1^{(n)}(0)$ with $n\ge2$ ~\cite{Majumdar12, PhysRevA.87.025803, Faraon08}, showing a quantum effect that a first excited photon favors the transmission of subsequent photons.  Without loss of generality, we adopt the criteria that $g_1^{(n)}(0)>1$ for $n=2,3,4$ to character the PIT effects~\cite{PhysRevLett.121.153601}. 

Regarding the experimental feasibility, our proposed scheme can be readily implemented using either single Rydberg atom~\cite{PhysRevLett.104.195302, madjarov2020high} or alkaline-earth-metal atom~\cite{PhysRevLett.117.220401, kolkowitz2017spin, PhysRevLett.118.263601, bromley2018dynamics} inside a high-finesse optical cavity by leveraging the advantages of their specific energy-level structures. Indeed, experimental setups for studying the generalized multiphoton Rabi model have already been realized or proposed in various experimental systems, including Rydberg atoms~\cite{PhysRevLett.88.143601,PhysRevLett.59.1899}, trapped-ion~\cite{PhysRevLett.76.1796}, superconducting circuit~\cite{PhysRevA.92.063830,PhysRevA.97.013851},
and quantum dots ~\cite{PhysRevB.73.125304,PhysRevLett.120.213901} in microwave cavities.

For the numerical simulations, we set the cavity decay rate $\kappa=2\pi \times 150$kHz is set as the energy unit,  the capability has been demonstrated in current atom-cavity experiments~\cite{leonard17, Leonard2017}.  The corresponding atomic spontaneous decay rate for the long-lived excited state is fixed at $\gamma/\kappa=0.1$~\cite{PhysRevLett.122.123604, PhysRevA.105.043711}. Without loss of generality, we focus on the atom-cavity resonance regime with $\Delta_a = 2\Delta_c$ and fix an experimentally accessible single atom-cavity coupling strength $g/\kappa=4$~\cite{PhysRevLett.118.133604}. Additionally, we assume that the cavity driving amplitude $\eta/\kappa \leq 1.5$, the atom pump field strength $\Omega/\kappa \leq 1.5$, and the Kerr parameter $\chi/\kappa \leq 10$, which are within the current experimental capabilities~\cite{PhysRevLett.118.223605, Kirchmair2013}.

\subsection{Emerged multi-PB under the cavity driving}

We first consider the quantum properties of the system under a weak cavity driving strength with $\eta/\kappa=0.1$, while maintaining a constant atom-pump coupling strength at $\Omega=0$. Figures~\ref{cavityPB}(a) and~\ref{cavityPB}(b) delineate the variation of the zero-time delay second-order correlation function $g_1^{(2)}(0)$ and the steady-state intracavity photon number $n_s$ across different cavity-light detunings $\Delta_c$, with Kerr nonlinearity $\chi/\kappa=0$ and 8, respectively. Remarkably, at the single-photon resonance ($\Delta_c=0$), both scenarios of $\chi/\kappa=0$ and $\chi/\kappa=8$ exhibit pronounced PB with $g_1^{(2)}(0) \sim 10^{-4}$ and a considerable intracavity photon number $n_s \sim 0.04$, as indicated by the energy spectrum analysis depicted in  Fig.\ref{scheme}(b) (red lines), which shows the emergence of an anharmonicity ladder within the proposed system. Additionally, the absence of Kerr nonlinearity yields a symmetric profile for $g_1^{(2)}(0)$ about $\Delta_c/g=0$. The introduction of $\chi$, however, disrupts this symmetry, attributing the asymmetric behavior of $g_1^{(2)}(0)$ to the frequency shift induced by Kerr nonlinearity in the second dressed splitting of the energy spectrum. Figure~\ref{cavityPB}(c) illustrates the dependence of optimal $g_{\rm{opt}}^{(2)}(0)$ and the corresponding optimal steady-state photon number $n_{s,\rm{opt}}$ on Kerr nonlinearity $\chi$ at single-photon resonance. With increasing $\chi$, a slight rise in $g_{\rm{opt}}^{(2)}(0)$ is observed, whereas $n_{s,\rm{opt}}$ remains virtually constant. Furthermore, Fig.~\ref{cavityPB}(d) presents the interval dependence of the second-order correlation function $g_1^{(2)}(\tau)$ with $\chi/\kappa=8$ at $\Delta_c/g=0$. It is evident that $g_1^{(2)}(0)<1$ and $g_1^{(2)}(0)<g_1^{(2)}(\tau)$, confirming that the output photons are sub-Poissonian and photon-antibunching properties. 

 \begin{figure}[ht]
\includegraphics[width=1\columnwidth]{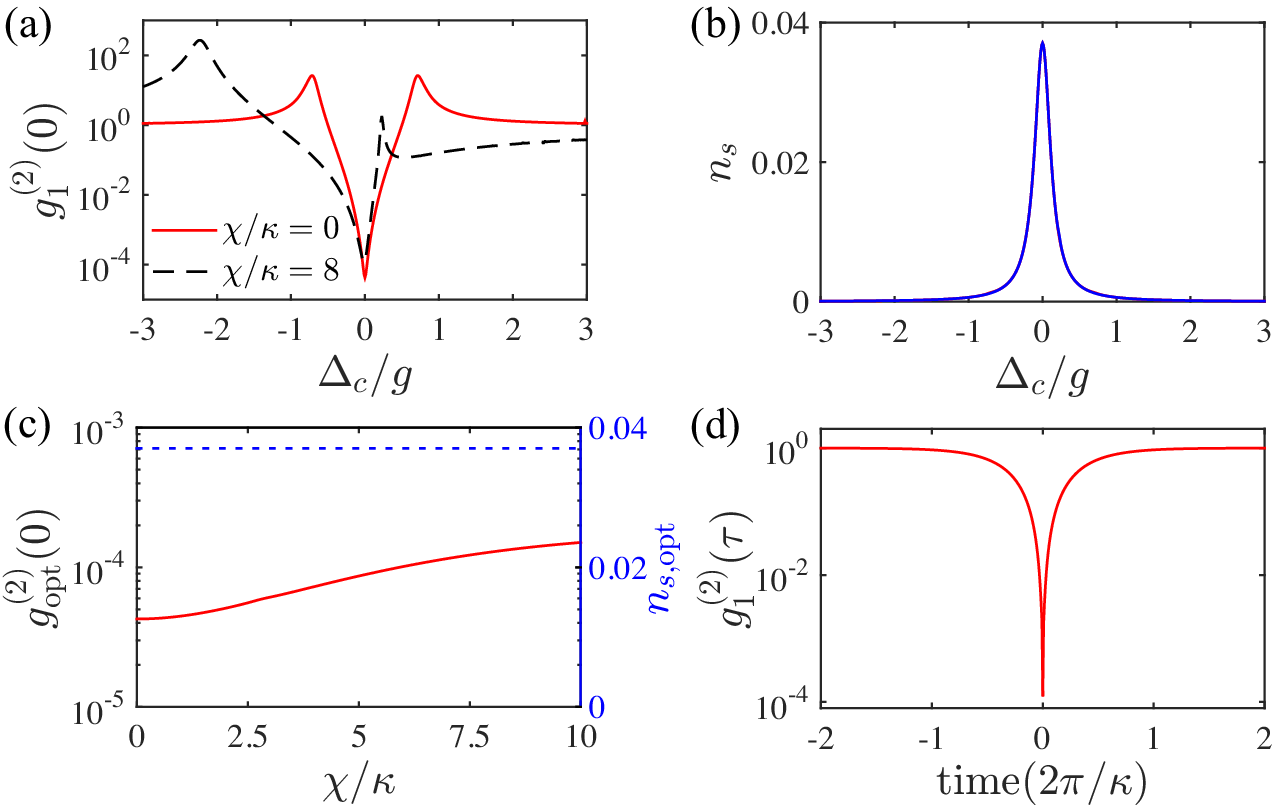}
\caption{(a) Second-order correlation function $g_1^{(2)}(0)$ and  (b) steady-state photon number $n_s$ as a function of cavity-light detuning $\Delta_c$ for $\chi/\kappa=0$ and 8, respectively. (c) $g_1^{(2)}(0)$ and $n_s$ vs Kerr nonlinearity $\chi$. (d) The time interval $\tau$ dependence of $g_1^{(2)}(\tau)$ for $\Delta_c/g=0$ and $\chi/\kappa=8$. In (a)-(d), the other parameters are held constant at  $\eta/\kappa=0.1$ and $\Omega/\kappa=0$.} \label{cavityPB}
\end{figure}%

We next examine the system's behavior under conditions exceeding the limitations of weak cavity driving, which is outside the typical configurations of interest in studying PB. Figures~\ref{cavityPB2}(a) and~\ref{cavityPB2}(b) depict the steady-state intracavity photon number $n_s$ and zero-time delay correlation functions $g_1^{(n)}(0)$ (for $n=2, 3, 4$) as a function of $\Delta_c$, under the conditions of $\eta=0.9$ and $\chi/\kappa=8$. Notably, single-PB manifests at the single-photon resonance with $g_1^{(2)}(0)=0.1$. In contrast to the weak driving scenario illustrated in Fig.~\ref{cavityPB}(b), two additional photon emission peaks are observed at two-photon resonances, corresponding to the red detuning $\Delta_c/g=-2.22$ and the blue detuning $\Delta_c/g=0.22$. Intriguingly, at the red detuning of $\Delta_c/g=-2.22$, the system exhibits three-PB, characterized by $g_1^{(2)}(0)=6.28$, $g_1^{(3)}(0)=1.18$, and $g_1^{(4)}(0)=8\times 10^{-2}$. This phenomenon illustrates the feasibility of transitioning between single-PB and three-PB by adjusting the cavity-light detuning $\Delta_c$. It is  noteworthy that increased cavity-driving strength induces three-PB at the two-photon resonance, a departure from the PIT effect observed in the absence of Kerr nonlinearity within the standard two-photon Jaynes-Cummings model~\cite{Tang:23}, indicating that the genesis of three-PB  is ascribed to the interplay of Kerr nonlinearity and the two-photon JC-type interaction.

 \begin{figure}[ht]
\includegraphics[width=1\columnwidth]{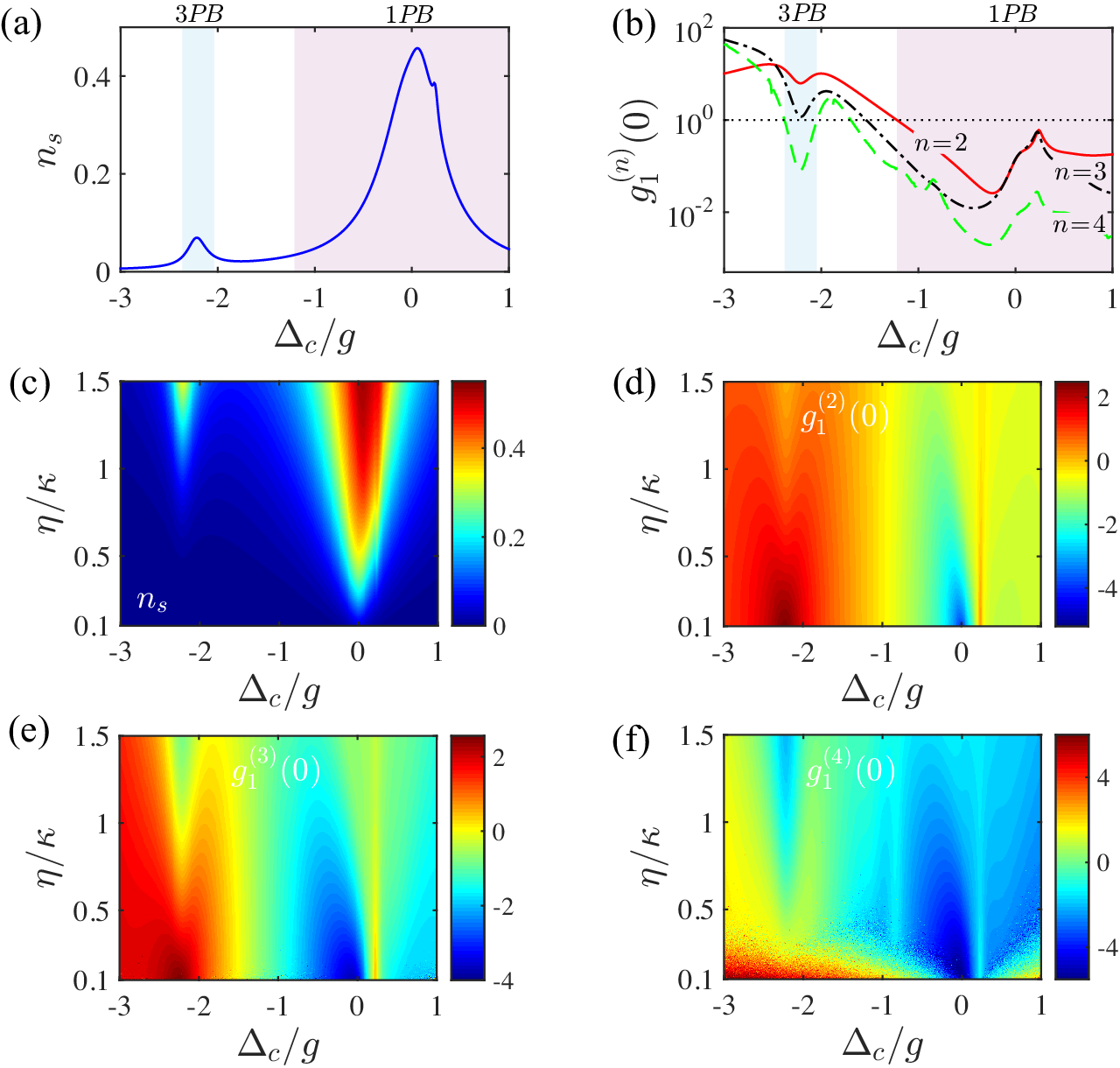}
\caption{(a) The steady-state photon number $n_s$ and (b) $n$th-order photon statistics $g_1^{(n)}(0)$ as a function of $\Delta_c$ for $\eta/\kappa=0.9$. The black dotted line in (b) indicating $g_1^{(n)}(0)=1$ is plotted to guide the eyes. Contour plots of $n_s$ (c), log [$g_1^{(2)}(0)$] (d), log [$g_1^{(3)}(0)$] (e), and log [$g_1^{(4)}(0)$] (f) on $\Delta_c-\eta$ parameter plane. The other parameters are $\chi/\kappa=8$, and $\Omega/\kappa=0$. } \label{cavityPB2}
\end{figure}%

To elucidate the multi-PB induced by Kerr nonlinearity in the two-photon JC model under cavity driving, Figs.~\ref{cavityPB2}(c)-3(f) present both the steady-state photon number $n_s$ and correlation functions $g_1^{(n)}(0)$ (for $n=2, 3, 4$) on a logarithmic scale, with $\chi/\kappa=8$,
highlighting the radiation emitted from the driven cavity and their quantum properties as functions of $\Delta_c$ and $\eta$, respectively. As can be seen, at low cavity-driving fields, a singular peak emerges at the single-photon resonance. As $\eta$ increases, two additional peaks materialize at two-photon resonances ($\Delta_c/g=-2.22$, $\Delta_c/g=0.22$), attributed to the augmented excitation intensity. Significantly, the correlation functions $g_1^{(n)}(0)$ exhibit a gradual decrease at the two-photon resonance $\Delta_c/g=-2.22$ with an increase in $\eta$, serving as critical indicators for identifying the emission of nonclassical quantum light. Conversely, an enhanced $g_1^{(2)}(0)$ at the single-photon resonance $\Delta_c=0$ with increasing $\eta$ signals a diminished single-PB effect.

Driven by these observations, Fig.\ref{cavityPB3}(a) outlines the optimal $g_{\rm{opt}}^{(2)}(0)$ and the corresponding optimal steady-state photon number $n_{s, \rm{opt}}$ at the single-photon resonance $\Delta_c=0$ against $\eta$. Notably, both $g_{\rm{opt}}^{(2)}(0)$ and $n_{s, \rm{opt}}$ exhibit a monotonous increase with the escalating cavity-driving strength $\eta$. This behavior is analogous to the standard JC model, as it is a result of the enhanced multiphoton transition processes and broadening of the energy spectrum under stronger excitation fields. Figure\ref{cavityPB3}(b) presents the optimal $g_{\rm{opt}}^{(n)}(0)$ and corresponding $n_{s, \rm{opt}}$ as a function of $\eta$ at the two-photon resonance $\Delta_c/g=-2.22$. Here, the values of $g_{\rm{opt}}^{(n)}(0)$ for $n=2, 3, 4$ decrease at varying rates with an increase in $\eta$. Specifically, within the range of $0.57 \leq \eta \leq 0.93$ (light-blue region), photon statistics reveal $g_{\rm{opt}}^{(2)}(0)>1$, $g_{\rm{opt}}^{(3)}(0)>1$, and $g_{\rm{opt}}^{(4)}(0)<1$, signifying a three-PB. For $\eta>0.93$ (light-yellow region), the conditions $g_{\text{opt}}^{(2)}(0)>1$ and $g_{\text{opt}}^{(3)}(0)<1$ are satisfied, denoting the  presence of a two-PB. The generation of the multi-PB is attributed to the Kerr nonlinearity-enhanced vacuum Rabi splitting at the $n$th dressed states, effectively preventing the system from absorbing the following photons when $n$ photons are already present in the cavity. Significantly, as the value of $\eta$ increases, the three-PB transitions to the two-PB, and conversely, a decrease in $\eta$ leads to a transition from two-PB to three-PB. Therefore, the switching between two-PB and three-PB can be achieved by modulating the cavity-driving strength. This flexibility holds potential for quantum information science applications, especially in manipulating light at the few-photon level~\cite{Giovannetti2011}.

 \begin{figure}[ht]
\includegraphics[width=1\columnwidth]{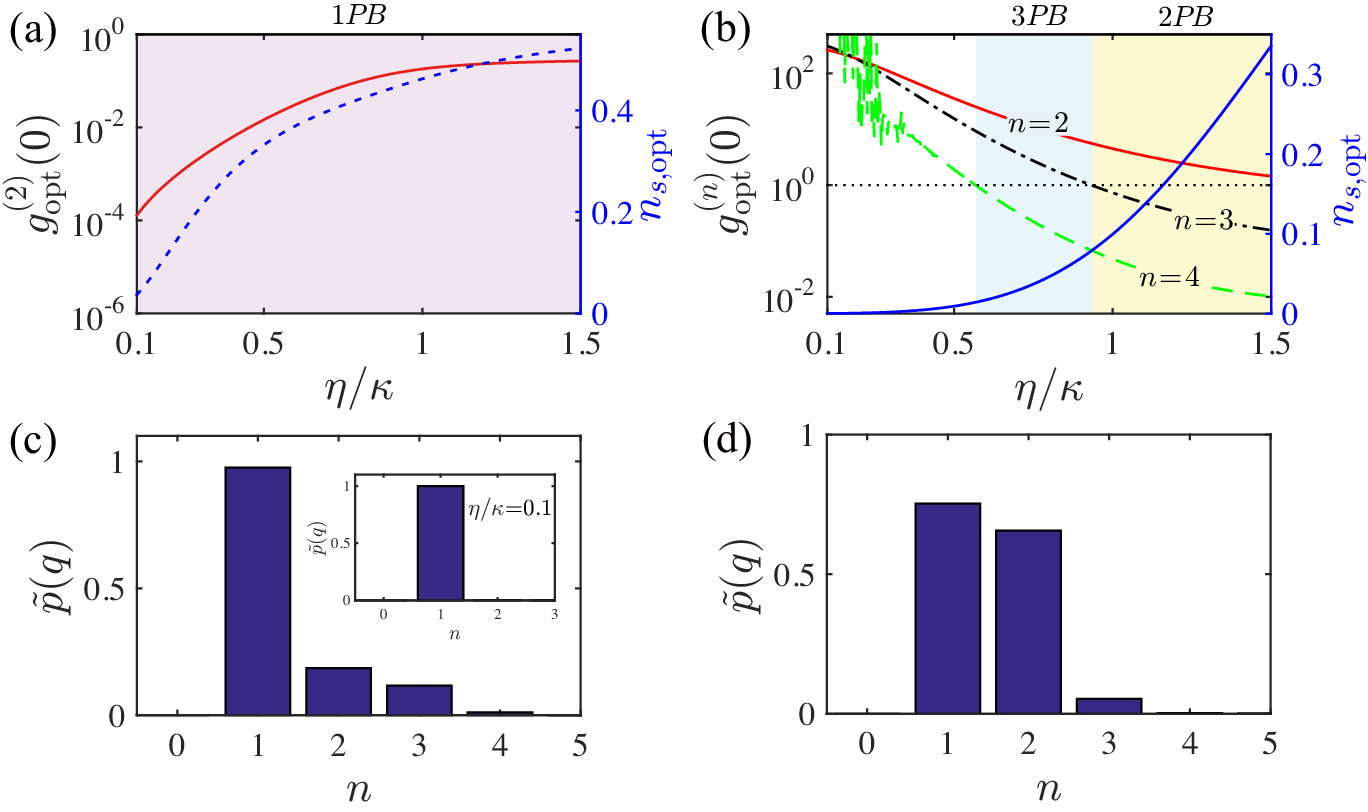}
\caption{[(a),(b)] depict the optimal $g^{(n)}_{\rm{opt}}(0)$ and the corresponding $n_{s, \rm{opt}}$ as a function of $\eta$ at single-photon resonance $\Delta_c/g=0$ and the two-photon resonance $\Delta_c/g=-2.22$, respectively. The black dotted lines in (b) indicating $g_1^{(n)}(0)=1$ is plotted to to provide visual guidance. (c) The amplitude of  the steady-state photon-number distribution $\tilde{p}(q)$ at the single-photon resonance with $\eta/\kappa=0.9$. The inset displays $\tilde{p}(q)$ with $\eta/\kappa=0.1$. (d) $\tilde{p}(q)$ at the two-photon resonance with $\eta/\kappa=0.9$. The other parameters are $\chi/\kappa=8$, and $\Omega/\kappa=0$. } \label{cavityPB3}
\end{figure}%

To further elucidate the photon statistics characteristics of the emitted photons, we examine the amplitude of the steady-state photon-number distribution, denoted by $\tilde{p}(q)=\sqrt{qp(q)/n_s}$, where $p(q)=\langle q | \hat{a}^\dagger\hat{a} |q \rangle$ indicates the probability of observing a $p$-photon Fock state. Figures~\ref{cavityPB3}(c) and~\ref{cavityPB3}(d) present $\tilde{p}(q)$ as a function of $q$-photon states. Specifically, Fig.~\ref{cavityPB3}(c) showcases $\tilde{p}(q)$ at single-photon resonance with $\eta/\kappa=0.9$. In comparison to the scenario under a lower cavity-driven field ($\eta/\kappa=0.1$) depicted in the inset, where nearly 100$\%$ single-photon nature is observed ($\tilde{p}(1) \approx 1$ and $\tilde{p}(q) \approx 0$ for $q\neq1$), an elevated cavity-driven field incites the emission of higher photon states, evidenced by $\tilde{p}(2) \approx 0.18$ and $\tilde{p}(3) \approx 0.12$ in Fig.~\ref{cavityPB3}(c). This phenomenon also explains the augmented $g_1^{(2)}(0)$ with the rise of $\eta$ at single-photon resonance. Figure~\ref{cavityPB3}(d) demonstrates $\tilde{p}(q)$ against $q$ at the two-photon resonance with $\eta/\kappa=0.9$. In the three-PB scenario, $\tilde{p}(q)$ dwindles to negligible levels for $q>3$, denoting the complete suppression of emission for $n\geq4$ photons.

\subsection{Transducer between two-photon bundles, three-photon bundles, and PIT}
We now consider exciting the atom directly using a classical field rather than the cavity-driving field with $\eta=0$. It is important to note that the effective coupling of the classical pump field to the atomic transition $|g\rangle   \leftrightarrow |m\rangle$ plays a crucial role in the  generation of multiphoton bundles emission. This process is analogous to the application of a strong two-photon parametric drive, represented by the interaction $\Omega(\hat{a}^{\dag 2} + \hat{a}^{2})$.

For an atom-pump strength $\Omega=0.65$, the mean cavity photon number  $n_s$ and the correlation functions  $g_1^{(n)}(0)$ ($n=2, 3$) vs the cavity-light detuning $\Delta_c$ with $\chi/\kappa=8$ are displayed in Figs.~\ref{atom}(a) and ~\ref{atom}(b) respectively. Notably, two peaks at the two-photon resonances ($\Delta_c/g=-2.22$ and $\Delta_c/g=0.22$), and a peak at the four-photon resonance ($\Delta_c/g=-0.85$) are observed. Particularly, at the red detuning of the two-photon resonance $\Delta_c/g=-2.22$, two-PB is achieved, characterizing by two-photon bunching ($g_1^{(2)}(0)=3.49$) and  three-photon antibunching [$g_1^{(3)}(0)=2.7\times10^{-2}$]. This two-PB phenomenon persists even when the detuning deviates from the two-photon resonance, as depicted by the light-yellow region in Fig.~\ref{atom}(b),  highlighting its potential in generating entangled photon pairs.  Remarkably, this two-PB effect exhibits a significant enhancement, nearly an order of magnitude stronger, in comparison to the conventional two-photon JC model without Kerr nonlinearity ($g_1^{(2)}(0)=1.17, g_1^{(3)}(0)=0.48$), attributable to the intensified vacuum Rabi splitting caused by the Kerr nonlinearity.

Concerning the four-photon resonance, photon statistics with $g_1^{(4)}(0)>g_1^{(3)}(0)>g_1^{(2)}(0)$ are observed, indicating the realization of PIT effect, as described by the light-green region in Fig.~\ref{atom}(b). This PIT effect arises from direct transition between the vacuum state to the second dressed states, which facilitates the subsequent photon absorption following the initial two-photon absorption. Intriguingly, at the blue detuning of the two-photon resonance ($\Delta_c/g=0.22$), we achieve three-PB, evidenced by two-photon and three-photon bunching ($g_1^{(2)}(0)=3.08$, $g_1^{(3)}(0)=1.04$) and four-photon antibunching ($g_1^{(4)}(0)=3\times10^{-2}$). The generation of three-PB can be attributed to the nearly degenerate vacuum Rabi splittings for $\Delta_{2+}$ and $\Delta_{3+}$, as illustrated in Fig.~\ref{scheme}(c).  This results in the emission of not only two-PB but also three-PB due to the anharmonic energy spectrum. Noteworthily, the phenomena of two-PB, PIT and three-PB occur in large parameter regions, as demarcated by the light-yellow, -green, -blue regions in Figs.~\ref{atom}(a) and ~\ref{atom}(b), which are conducive to the experimental realization of these nonclassical states. Additionally, a quantum three-phase transducer enabling transition between  two-PB, PIT and three-PB is achieved by tuning the cavity-light detuning. This capability opens up exciting possibilities for controlling and tailoring the photon statistics in quantum systems, with potential applications in quantum communication and emerging quantum technologies~\cite{Brien09, BLIOKH20151}.

\begin{figure}[ht]
\includegraphics[width=1\columnwidth]{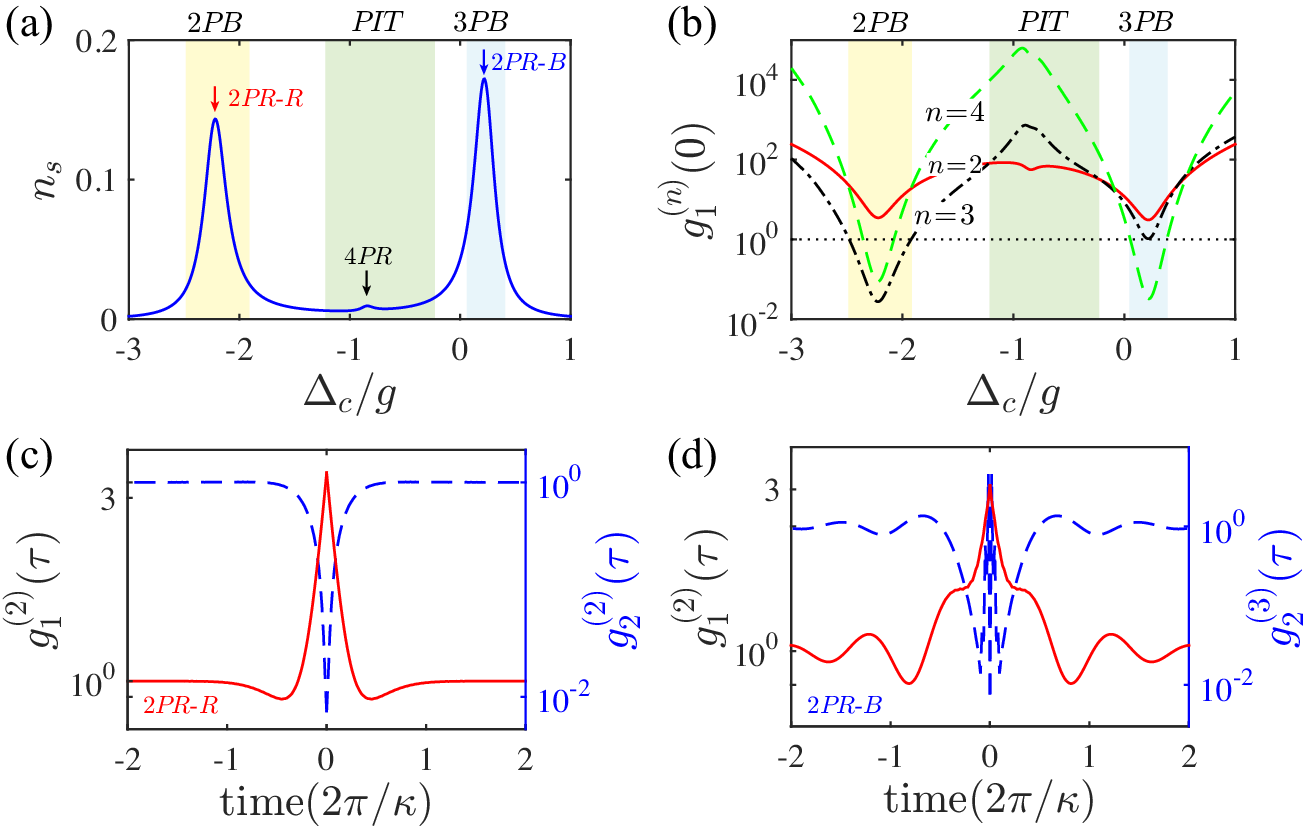}
\caption {(a) Intracavity photon number $n_s$ and (b) correlation functions $g_1^{(n)}(0)$ vs $\Delta_c$. In (b), the black dotted line indicating $g_1^{(n)}(0)=1$ is included as a visual reference. (c) Distribution of $g_1^{(2)}(\tau)$ (red solid line) and $g^{(2)}_2(\tau)$ (blue dashed line) at the two-photon resonance with $\Delta_c/g=-2.22$. (d) Distribution of $g_1^{(2)}(\tau)$ (red solid line) and $g^{(3)}_2(\tau)$ (blue dashed line) at the two-photon resonance with $\Delta_c/g=0.22$. Here, the $n$PR represents $n$-photon resonance and $n$PR-R (B) denotes the red  (blue) detuning of the $n$-photon resonance for brevity. The other parameters are $\chi/\kappa=8$, $\Omega/\kappa=0.65$, and $\eta/\kappa=0$. } \label{atom}
\end{figure}%

Figure~\ref{atom}(c) delineates the interval dependence of correlation functions $g_1^{(2)}(\tau)$ (red line) and $g_2^{(2)}(\tau)$ (blue-dashed line) at the red detuning of the two-photon resonance ($\Delta_c/g=-2.22$). A peak in $g_1^{(2)}(\tau)$ while a dip in $g_2^{(2)}(\tau)$ are observed, suggesting $g_1^{(2)}(\tau)<g_1^{(2)}(0)$ and $g_2^{(2)}(\tau)>g_2^{(2)}(0)$ respectively,  thereby confirming the occurrence of a two-photon bundles state. Furthermore, the decay time for $g_1^{(2)}(\tau)$ and $g_2^{(2)}(\tau)$  is nearly identical and proportional to  $1/\kappa$,  indicating a consistent decay timescale for both single-photon bunching and separated photon-bundle antibunching within the two-photon bundles state.
Analogously, the three-photon emission character at the blue detuning of two-photon resonance ($\Delta_c/g=0.22$) unequivocally satisfies the criteria for a three-photon bundles state, as illustrated in Fig.~\ref{atom}(d), where $g_1^{(2)}(\tau)<g_1^{(2)}(0)$ and $g_2^{(3)}(\tau)>g_2^{(3)}(0)$. 
Contrasting with $n$PB, $n$-photon bundles release their energy in $n$-photon groups in nature, exhibiting photon bunching for individual photons and antibunching for separated photon-bundles. This unique emission profile holds great promise for applications such as $n$-photon lasers and photon guns~\cite{munoz2014emitters}, novel quantum light sources and quantum metrology~\cite{walther2006cavity, giovannetti2004quantum}.    

\begin{figure}[ht]  
\includegraphics[width=0.8\columnwidth]{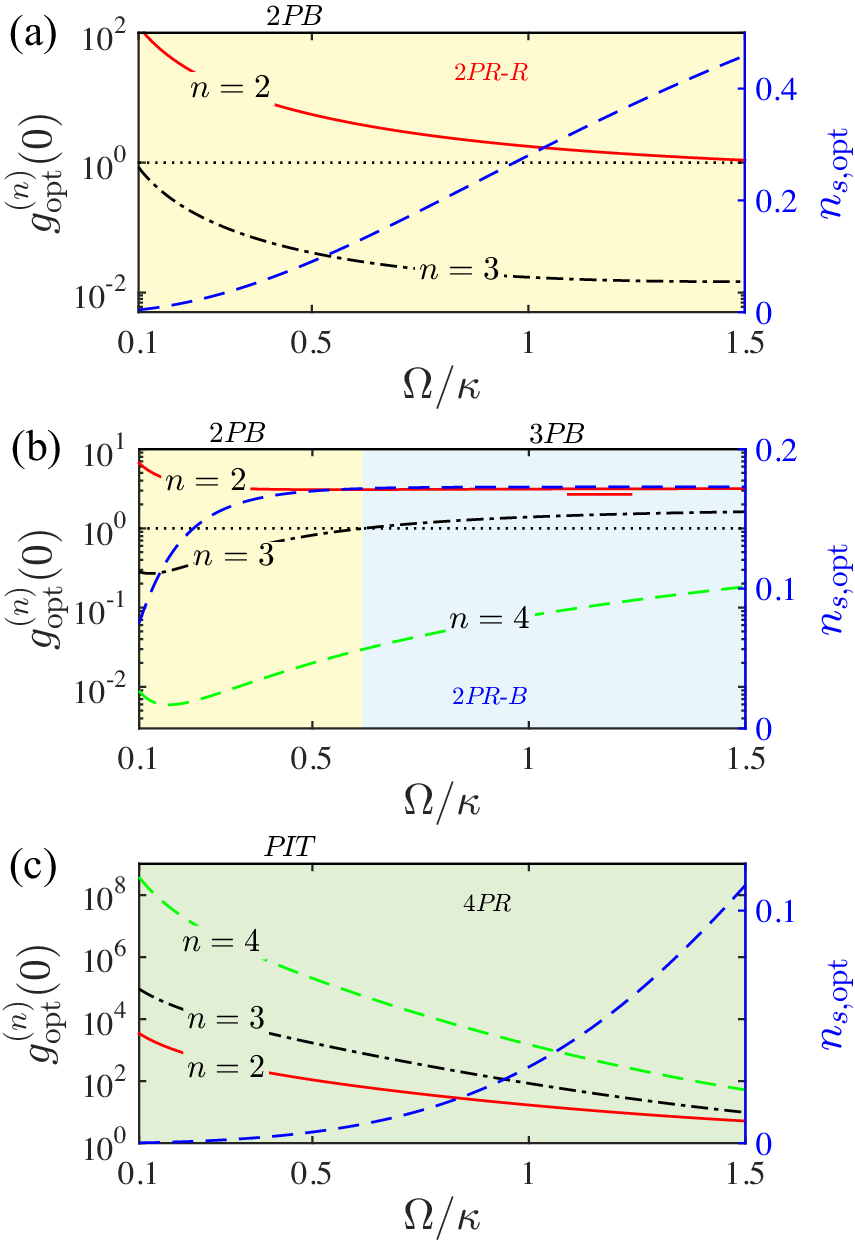}
\caption{(a)-(c) $\Omega$ dependence of optimal correlation functions $g_{\rm{opt}}^{(n)}(0)$ and the corresponding steady-state photon number $n_{s, \rm{opt}}$ for $\Delta_c/g=-2.22$, $\Delta_c/g=0.22$, and $\Delta_c/g=-0.84$, respectively. The black dotted lines indicating $g_{\rm{opt}}^{(n)}(0)=1$ are plotted  to provide visual guidance. The other parameters are $\chi/\kappa=8$, and $\eta/\kappa=0$.} \label{atom1} 
\end{figure}%

In Fig.~\ref{atom1},  the optimal correlation functions $g_{\rm{opt}}^{(n)}(0)$ ($n=2,3,4$) and the corresponding $n_{s, \rm{opt}}$ are plotted as a function of the atom-pump strength $\Omega$, for both the two-photon and four-photon resonances, with a Kerr nonlinearity of $\chi/\kappa=8$. At the red detuning of the two-photon resonance [Fig.~\ref{atom1}(a)], we observe that $g_{\rm{opt}}^{(2)}(0)>1$ and $g_{\rm{opt}}^{(3)}(0)<1$ within the range of  the displayed $\Omega$ values (light-yellow region), signifying a pronounced two-PB phenomenon. As the atom-pump strength $\Omega$ increases, the values of the correlation functions $g_{\rm{opt}}^{(n)}(0)$ ($n=2,3$) decrease monotonically, indicating an enhanced two-PB effect. In fact, at very low atom-pump strengths, the excited average photon number is less than two photons, making it challenging to achieve the two-photon transition. With an increase in $\Omega$, the average photon numbers grow, facilitating the generation of high-quality two-PB. However, at a sufficiently high atom-pump strength, multiphoton transitions occur due to the significantly increased average photon numbers, which in turn disrupts the two-PB. 

Figure~\ref{atom1}(b) elucidates the $\Omega$-dependent behavior of $g_{\rm{opt}}^{(n)}(0)$ ($n=2,3,4$) and $n_{s, \rm{opt}}$ at the blue detuning of the two-photon resonance, under identical system parameters as in Fig.~\ref{atom1}(a).  Here, an initial decline followed by an increment in the values of $g_{\rm{opt}}^{(n)}(0)$ ($n=2,3,4$) with rising $\Omega$ is observed, unveiling a local optimum for two-PB at $\Omega/\kappa=0.12$. Notably, a transition to three-PB is marked by $g_{\rm{opt}}^{(3)}(0)$ surpassing 1 (highlighted in light-blue). Owing to the nearly degenerate vacuum Rabi splittings $\Delta_{2, +} $ and $\Delta_{3, +}$, as illustrated in Fig.~\ref{scheme}(c), a strong atom-pump strength favors the generation of a three-photon transition through the heightened average photon numbers. Remarkably, by increasing the atom-pump strength up to $\Omega/\kappa=0.62$, the two-PB effect can be transformed into three-PB, and vice versa, signifying the achievement of an optical transducer between two-PB and three-PB states.

Figure~\ref{atom1}(c) depicts the photon statistics vs $\Omega$ at the four-photon resonance, where the PIT effect is discernible across the entire $\Omega$ parameter range, as denoted by the light-green region, despite a decrement in $g_{\rm{opt}}^{(n)}(0)$ ($n=2,3,4$) with an upsurge in $\Omega$. These findings underscore the potential of effectuating an optical transducer between two-PB, three-PB and PIT by maneuvering the cavity-light detuning $\Delta_c$, with $\Omega/\kappa\geq 0.62$. This capability has significant implications for applications in quantum communication and emerging quantum technologies.

\begin{figure}[ht]
\includegraphics[width=1\columnwidth]{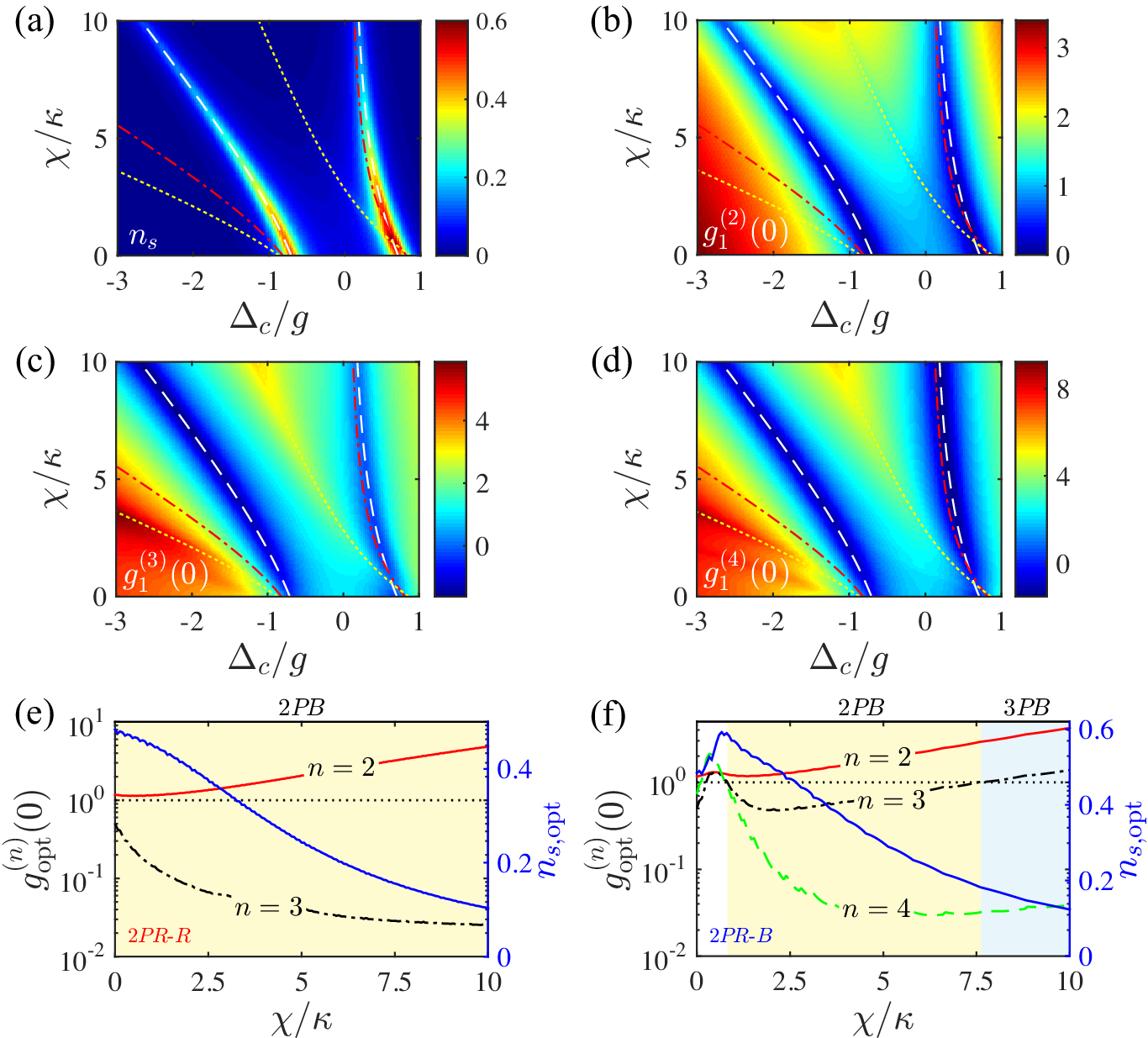}
\caption{(a) Contour plots illustrating the dependencies of  (a) $n_s$,  (b) log [$g_1^{(2)}(0)$],  (c) log [$g_1^{(3)}(0)$] , and  (d) log [$g_1^{(4)}(0)$]  on both the cavity-light detuning  $\Delta_c$ and Kerr nonlinearity $\chi$.   Panels (e) and (f) display the values of $g_{\rm{opt}}^{(n)}(0)$ and the corresponding $n_{s, \rm{opt}}$ as a function of Kerr nonlinearity $\chi$ for two-photon resonances $\Delta_c/g=-2.22$ and $\Delta_c/g=0.22$, respectively. In panels (a)-(d), the white, red, and yellow lines represent the analytical energy spectrum calculated using Eq.~(\ref{Energy}). Dotted black lines in panels (e) and (f) indicate  $g_{\rm{opt}}^{(n)}(0)=1$ for visual reference. The other parameters are set as $g/\kappa=4$, $\Omega/\kappa=0.65$, and $\eta/\kappa=0$.} \label{Kerr}
\end{figure}%

Next, we examine the influence of Kerr nonlinearity on the photon statistics of the system, holding the atom-pump strength at $\Omega/\kappa=0.65$. Figures~\ref{Kerr}(a)-\ref{Kerr}(d) display the intracavity photon numbers $n_s$ and the photon correlations $g_{\rm{opt}}^{(n)}(0)$ ($n=2,3,4$) on a logarithmic scale, highlighting the regime of the generated nonclassical states as a function of $\Delta_c$ and $\chi$. It is  apparent that the $n$-photon resonance ($n=2,3,4$) undergoes significant nonlinear shifts with increasing $\chi$, aligning closely with the analytically derived asymmetric energy splittings, denoted by the white dashed lines [$\Delta_{2,\pm}$], red dashed-dotted lines [$\Delta_{3,\pm}$], and yellow dotted lines [$\Delta_{4,\pm}$]. Here, we focus on the sing-PB and multi-PB effect in the system. 

To intuitively understand the photon statistics, we illustrate the optimal correlation functions $g_{\rm{opt}}^{(n)}(0)$ and the respective $n_{s, \rm{opt}}$ as a function of $\chi$ at two-photon resonances $\Delta_c/g=-2.22$ and $\Delta_c/g=0.22$ in Figs.\ref{Kerr}(e) and \ref{Kerr}(f), respectively. In Fig.\ref{Kerr}(e), at the red detuning of two-photon resonance $\Delta_c/g=-2.22$,  two-PB are achieved with two-photon bunching and three-photon antibunching throughout the entire parameter space (light-yellow region). Moreover, as $\chi$ increases, the values of $g_{\rm{opt}}^{(2)}(0)$ rise while those of $g_{\rm{opt}}^{(3)}(0)$ fall, accompanied by a decline in photon numbers $n_{s, \rm{opt}}$, indicating an intensified two-PB effect. The enhanced two-PB effect can be ascribed to the strong anharmonic energy spectrum brought about by the Kerr nonlinearity, which facilitates the two-photon transition while inhibiting $m$-photon transitions ($m>2$).

Similarly, Fig.\ref{Kerr}(f) maps out $g_{\rm{opt}}^{(n)}(0)$ ($n=2,3,4$) and the corresponding $n_{s, \rm{opt}}$ at the blue detuning of two-photon resonance $\Delta_c/g=0.22$ as a function of $\chi$. At $\chi/\kappa=0$, the values of $g_{\rm{opt}}^{(n)}(0)$ ($n=2,3,4$) and $n_{s, \rm{opt}}$ coincide at both blue and red detuning of the two-photon resonances, in agreement with outcomes from the standard two-photon Jaynes-Cummings model devoid of Kerr nonlinearity\cite{Tang:23}. Nevertheless, with a minimal $\chi$, specifically $0<\chi/\kappa \leq0.82$, the values of $g_{\rm{opt}}^{(n)}(0)$ ($n=2,3,4$) and $n_{s, \rm{opt}}$ initially rise then diminish as $\chi$ grows. This behavior arises from the influence of the four-photon transition at the four-photon resonance, which can be deduced from the behavior of the $n$th vacuum Rabi splittings as depicted in Fig.~\ref{scheme}(c). For $0.82<\chi/\kappa \leq7.6$, the emission photons exhibit a two-PB effect, where $g_{\rm{opt}}^{(2)}(0)>0$ and $g_{\rm{opt}}^{(3)}(0)<0$ (the light-yellow region). Specifically, $g_{\rm{opt}}^{(3)}(0)$ displays a dip feature, signifying an optimal $\chi$ for two-PB generation. For $\chi/\kappa>7.6$, where $\Delta_{2+}$ and $\Delta_{3+}$ nearly converge, the values of $g_{\rm{opt}}^{(n)}(0)>0$ ($n=2,3$) and $g_{\rm{opt}}^{(4)}(0)<0$, indicating a shift into the three-PB regime (the light-blue region). This shift is attributed to the strong interaction favoring multiphoton transitions from lower to higher energy levels. Remarkably, $g_{\rm{opt}}^{(4)}(0)$ experiences a slight increase in this regime, stemming from the competition between the energy spectrum anharmonicity induced by Kerr nonlinearity $\chi$ and the atom-cavity coupling strength $g$. This findings suggests that an optical transducer between 2PB and 3PB can also be realized by adjusting the Kerr nonlinearity.

\subsection{Enhancement of three-PB via combined driving fields}

In this section, we delve into the emergence of rich nonclassical quantum states within the framework of our system, under the influence of simultaneous driving by both cavity and atomic fields. Set the atom-pump field at $\Omega/\kappa=0.65$ alongside a cavity-driving field of $\eta/\kappa=0.35$, we observe a range of intriguing quantum states at the $n$-photon resonances, as displayed in Figs.~\ref{atom_cavity}(a) and~\ref{atom_cavity}(b). Specifically, our analysis reveals the presence of two- and three-PB at the two-photon resonances, PIT at the four-photon resonance, and single-PB at the single-photon resonance. Noteworthy is the significant amplification of the three-PB at the red detuning of the two-photon resonance ($\Delta_c/g=-2.22$), achieved through the synergistic effect of a stronger cavity-driven field coupled with an atom-pump field.

\begin{figure}[ht]
\includegraphics[width=1\columnwidth]{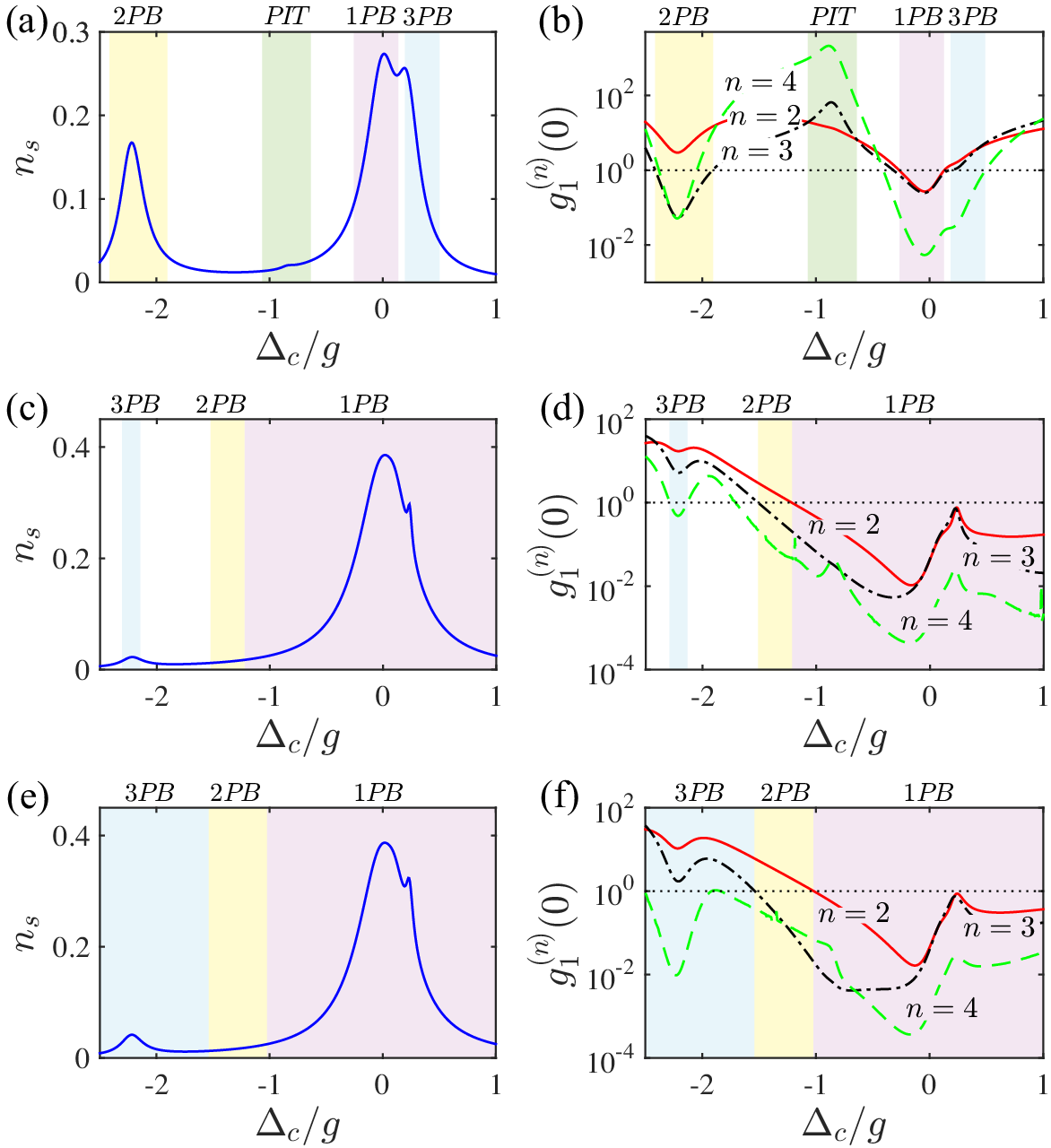}
\caption{(color online). (a) $n_s$ and (b) $g_1^{(n)}(0)$ as a function of $\Delta_c$ with $\Omega/\kappa=0.65$ and $\eta/\kappa=0.35$. $n_s$ and $g_1^{(n)}(0)$ as a function of $\Delta_c$ with $\eta/\kappa=0.65$ for $\Omega/\kappa=0$ [(c) and (d)] and $\Omega/\kappa=0.1$ [(e) and (f)], respectively. The other parameters are $\chi/\kappa=8$. The black dotted lines indicating $g_1^{(n)}(0)=1$ are plotted to guide the eyes.} \label{atom_cavity}
\end{figure}%

For illustrative purposes, we present the intracavity photon number $n_s$ and correlation functions $g_1^{(n)}(0)$ for $n=2, 3, 4$ as a function of cavity-light detuning $\Delta_c$ with fixing driving field for different atom-pump field. Figures~\ref{atom_cavity}(c) and~\ref{atom_cavity}(d) showcase the scenario absent an atom-pump field ($\Omega/\kappa=0$) with a stronger cavity-driven field strength ($\eta/\kappa=0.65$). In this setting, a three-PB emerges at $\Delta_c/g=-2.22$, characterized by the correlation function values of $g_1^{(2)}(0)=17.23$, $g_1^{(3)}(0)=5.26$, $g_1^{(4)}(0)=0.5$, and an intracavity photon number of $n_s=0.02$. Introducing a minimal atom-pump field strength ($\Omega/\kappa=0.1$), as depicted in Figs.~\ref{atom_cavity}(e) and~\ref{atom_cavity}(f), results in a marked improvement in the three-PB at the same detuning  $\Delta_c/g=-2.22$ ($g_1^{(2)}(0)=10.54$, $g_1^{(3)}(0)=1.73$, $g_1^{(4)}(0)=9.6\times10^{-3}$ and $n_s=0.04$), since $g_1^{(4)}(0)$ has a more than 52-fold reduction compared to the scenario devoid of the atom-pump field. 

Our findings underscore the potential to manipulate  a wide array of nonclassical states and phenomena through precise adjustments of the cavity-light detuning, atom-pump field strength, and cavity-driven field intensity. This versatility makes it a versatile platform for quantum optics experiments and potential applications in quantum computing and quantum simulation~\cite{Bennett00, Georgescu14}.

\section{conclusions}

In conclusion, we explore  the generation of nonclassical light emission within a Kerr-type two-photon JC model via analytically the energy spectrum and numerically calculating the master equation for a moderate atom-cavity coupling. A notable amplification in vacuum Rabi splitting for the $n$th manifold is observed with modulating the Kerr nonlinearity, corresponding to the well-resolved multiphoton resonance with significantly enhanced energy-spectrum anharmonicity, which culminates in the emergence of rich quantum lights emission in the system. For a large cavity-driven field, three-PB and two-PB can occur and convert into each other by adjusting the cavity-driven strength, with the system acting as a transducer between these states and single-PB by tuning the cavity-light detuning. When the system operates in atom-pump mode, the emission of two- and three-photon bundles can be achieved by adjusting the atom-pump strength or Kerr nonlinearity, enabled by near-degenerate cavity-light detuning at two- and three-photon resonances. Moreover, the combined operation of cavity-driven and atom-pump fields reveals a tunable transducer among two-PB, PB, PIT, and three-PB, with significant enhancement of three-PB at the red detuning of two-photon resonance through the atom-pump strength control. Our scheme for generating diverse PB phenomena, PIT, and related quantum states holds promising prospects for application in quantum information science and technology, including  entangled photon sources and adaptable quantum devices~\cite{PhysRevLett.129.193604}, unveiling the potential  in paving new avenues for quantum technology development.

\section{ACKNOWLEDGMENT}\label{acknow}

This work was supported by the National Natural Science Foundation of China (Grant No.12374365, No. 12274473, and No. 12135018) and the Fundamental Research Funds for the Central Universities (Grant No. 24qnpy120).

%
%

\end{document}